# Poly(methyl methacrylate) - Palladium clusters nanocomposite formation by supersonic cluster beam deposition: a method for microstructured metallization of polymer surfaces

Luca Ravagnan, Giorgio Divitini, Sara Rebasti[1], Mattia Marelli, Paolo Piseri and Paolo Milani[2]

CIMAINA and Dipartimento di Fisica, Università di Milano, Via Celoria 16, I-20133 Milano, Italy

E-mail: pmilani@mi.infn.it

**Abstract**

Nanocomposite films were fabricated by supersonic cluster beam deposition (SCBD) of palladium clusters on Poly(methyl methacrylate) (PMMA) surfaces. The evolution of the electrical conductance with cluster coverage and microscopy analysis show that Pd clusters are implanted in the polymer and form a continuous layer extending for several tens of nanometers beneath the polymer surface. This allows the deposition, using stencil masks, of cluster-assembled Pd microstructures on PMMA showing a remarkably high adhesion compared to metallic films obtained by thermal evaporation. These results suggest that SCBD is a promising tool for the fabrication of metallic microstructures on flexible polymeric substrates.

---

[1] present address: STMicroelectronics, Agrate Brianza (MI), Italy
[2] Author to whom correspondence should be addressed

The interest for micro- and nanomanufacturing of polymeric materials is continuously increasing driven by different fields such as flexible optoelectronics and microfluidics for biomedical and chemical analysis systems [1]. The need of polymer-based microdevices incorporating catalytic, sensing, signal conditioning and actuating functions require the ability to integrate on polymer substrates metallic nanoparticles with controlled dimensions and density in order to create active layers, contacts, wires, circuits and interconnections [1, 2]. Metal-polymer nanocomposites can find application in the production of nonlinear optical systems [3-5], magnetic devices [5-7], strain-gauges [8] and antibacterial coatings [5].

The fabrication of polymer-metal nanocomposites and the selective metallization of polymer surfaces have been obtained with various techniques: magnetron sputtering [9], ion implantation [3, 6, 10, 11], photo- or electron beam lithography [2], photoreduction [12], soft lithography and microcontact printing [13, 14]. Many of these methods are based on the insertion of atomic of molecular precursors in a pre-existing or co-deposited polymeric matrix and on the induced condensation of nanoparticles by physical or chemical stimuli. The independent control of the position, density and dimension of nanoparticles without modifying or damaging the polymeric matrix remains a major problem [10]. A nanoparticle-based approach can represent a significant improvement compared to polymer metallization based on atomic physical vapour deposition of noble metals; this method, although cheap and easily scalable, has poor performances in terms of layer adhesion and attainable lateral resolution [15].

Poly(methyl methacrylate) (PMMA) is a polymer widely used for microfluidics and optoelectronic applications because of its mechanical (glass transition temperature of 124°C) and chemical properties [16, 17]. The incorporation of noble metal nanoparticles in PMMA is used to control its optical properties [18] while metallization of PMMA is used for electrophoresis [17, 19] and dielectrophoresis on microfabricated devices [20].

Here we report about the formation of metal-polymer nanocomposites by supersonic cluster beam deposition (SCBD) of palladium nanoparticles on PMMA. Using stencil masks we directly fabricated cluster-assembled metallic microstructures on PMMA and we characterized their adhesion and stability. The evolution of the electrical conductivity of the nanocomposite as a function of cluster coverage was characterized in situ during the deposition.

Pd clusters were produced by a pulsed microplasma cluster source (PMCS) as described in detail elsewhere [21]. A palladium target rod is placed in the PMCS ceramic cavity where a solenoid valve delivers pulses of He gas. A He plasma is ignited by a pulsed electric discharge between the Pd rod (cathode) and an anode, producing the ablation of the target. The ablated Pd atoms are thermalized inside the cavity by collision with the inert gas condensing into clusters, the He-cluster mixture is then expanded in a vacuum chamber through a nozzle to form a supersonic beam. Typical cluster size, as produced by the PMCS, is in the range from few tents to several thousands atoms per cluster with a log-normal distribution peaked at few hundreds atoms per cluster, the kinetic energy upon landing is on the order of 0.5 eV per atom. Aerodynamic focussing techniques are employed to obtain a highly collimated cluster beam (with a divergence below 50 mrad) [22]. The deposition takes place on substrates intercepting the supersonic beam in a second differentially pumped chamber separated from the expansion chamber by an electroformed skimmer.

Pd clusters were deposited on two different substrates: PMMA (Goodfellow) (0.5 mm thickness) and MgO (100); on both substrates two gold electrodes (1 µm thickness) separated by a 1 mm gap were previously evaporated to provide electrical contact with the growing layer of deposited clusters. The substrates were mounted on a variable temperature sample holder allowing the collection of a portion of the cluster beam by a quartz microbalance in order to continuously monitor the cluster deposition during the electrical measurements (typical deposition rate 0.5 nm/min, corresponding to a cluster flux of $2 \times 10^{10}$ cluster/s cm$^2$) [23]. The amount of deposited clusters is expressed in terms of the mean film thickness measured by the quartz microbalance. The microbalance output was accurately calibrated by measuring with an atomic force microscopy (AFM) the thickness of a set of reference samples deposited on MgO substrates. The roughness of the substrates (0.2 ± 0.1 nm for MgO and 1.4 ± 0.1 nm for PMMA) and the average size of the clusters (i.e. their diameters, measured as the heights of the isolated nanoparticles on MgO deposited by a single shot of the PMCS) were also measured by AFM. The current flowing in the cluster-assembled films was measured by a Keithley 6517 electrometer ($10^{-14}$ A sensitivity) at a constant applied voltage of 1.5 V between the gold electrodes. MgO was used as a reference, since the organization and growth of Pd nanostructures and films on MgO (100) has been extensively studied [24].

Figure 1 shows the conductance evolution with increasing amount of deposited clusters for depositions on PMMA and on MgO at room temperature and for PMMA heated to 95°C; for both substrates the characteristic evolution of conductance across a percolation threshold is clearly observed at room temperature [25]. Below a critical coverage the deposits have an insulating behaviour since the deposited Pd clusters do not form metallic paths connecting the two electrodes; beyond the percolation threshold, the addition of new clusters causes a rapid increase of the conductance by 8 orders of magnitude.

The conductance evolution observed for MgO and PMMA at room temperature is qualitatively similar but different values are registered for the percolation threshold: ~1 nm on MgO against ~5 nm on PMMA. Results from conductance evolution for the deposition performed on a PMMA substrate heated at 95°C show that the percolation threshold increases to 10 nm thickness and the conductance grows much slower after this threshold (figure 1).

Since the average size of the deposited clusters as determined by AFM is 2.7 ± 1.4 nm, the threshold observed on MgO is compatible with a ballistic deposition of clusters onto the flat surface without substantial reorganization of the growing layer. A percolation threshold at about 50% of substrate coverage is typical for the growth a 2D random network [25]. Significantly larger values for the percolation threshold, as observed here in the case of PMMA, can be observed in the presence of high mobility and cohesive energy of the deposited species thus causing their aggregation as 3D islands [25]. This interpretation of the observed percolation threshold values would imply that Pd clusters have a significantly larger mobility on PMMA than on MgO (100), which is at odd with observations reported by several authors [2, 24].

An alternative explanation of the observed behaviour can be given by considering the possibility that the Pd clusters do not remain on the polymer surface but penetrate inside the polymer matrix. In this case the percolating system is 3D instead of 2D and the amount of deposited material does not span surface coverage but the density of clusters embedded inside the polymer matrix. The percolation threshold should be then expressed in terms of a volume fraction whose relation to the thickness measured by the quartz microbalance depends on the penetration depth of the clusters in the polymer matrix and on their aggregation.

In figure 2 transmission electron microscopy (TEM) micrographs of cross sections of the Pd-PMMA samples deposited on substrates kept at room temperature and at 95°C are shown. At both temperatures, for a nominal film thickness lower than 1 nm (we define nominal film thickness as the thickness of the film produced by the same amount of clusters deposited on a MgO substrate), individual metal nanoparticles and nanoparticle agglomerates are distinguishable beneath the polymer surface occupying a well-defined region reaching a depth of 50 nm at RT and 70 nm at 95°C (figure 2a and 2b) due to the penetration of the clusters beneath the polymer surface. This observation provides direct support to the interpretation of electrical transport measurements in terms of a 3D percolation at both temperatures. The result is remarkably surprising as diffusion of clusters in polymers is reported only above the glass transition temperature $T_g$ (over 120°C in the case of PMMA) [26], while only diffusion of isolated noble metal atoms [2] has been observed for substrate temperatures below $T_g$. The cluster distribution inside the polymer does not show the typical exponential decrease in concentration with depth, nor a decrease of the mean cluster size; the dispersion is uniform and very different from what is typically observed in the case of clusters formed by thermal-induced condensation of deposited isolated atomic species [26, 27]. By increasing the nominal thickness, the buried clusters form a more compact layer which starts to increase in thickness surfacing to the PMMA surface and forming a continuous film firmly anchored to the polymer substrate (figure 2c and 2d).

This cluster distribution in the polymer is quite similar to what typically observed for ion implantation in polymers [4, 10]. In this case the penetration depth is dependent on the kinetic energy of the ions and, as a consequence, the post-implantation thermal annealing causes the formation of nanoparticles in a well-defined region beneath the polymer surface. On this basis we infer a role of cluster kinetic energy for the penetration of Pd clusters in PMMA. The kinetic energy acquired by a Pd cluster in the supersonic expansion is of the order of 0.5 eV/atom (velocity of 1000 m/s); this is significantly lower compared to kinetic energies typical of monomer ion implantation used for the production of polymer nanocomposites [4, 10, 28]. A mechanism capable of explaining a substantial increase of the penetration depth of clusters in ion implantation processes as compared to monomers, the so-called 'clearing-the-way' effect [29], has been proposed and shown to be relevant for moderate energy cluster implantation in soft van-der-Waals materials [30]. According to this model the clusters collide with the substrate atoms transferring them sufficient momentum to clear the way for the penetration of the particle: this becomes more efficient as the cluster size increases [30]. In our case this mechanism cannot account for a penetration depth of 50-70 nm for Pd clusters. Nevertheless, apart from a purely momentum driven effect, one has also to consider that the cluster impact induces locally shock conditions that increase remarkably the pressure and temperature of the polymer in the impact area [31]. Thus, even if the heating of the whole polymer surface during the cluster deposition is negligible (the cluster-beam mean power density is $1 \times 10^{-6}$ W/cm$^2$, much lower than the typical 0.3 W/cm$^2$ encountered with monomer implantation [28]), locally the properties of the polymer are radically changed by the impact, and this may be at the base of the very high penetration depth observed.

We have also explored the possibility of using SCBD for the fabrication of metallic microstructures on polymer surfaces with superior adhesion and stability properties. By exploiting the favorable characteristics of supersonic cluster beams for microstencil lithography [32], we have deposited a palladium microwire (2 mm long, 10 μm wide and 50 nm thick) on PMMA heated to 95°C using a stencil mask (figure 3a). In figure 3b we show the microwire after having performed a Scotch® Tape Test [33]: a Magic Scotch tape (2.4 cm width, produced by 3M) was firmly attached to the substrate at room temperature by applying a gentle pressure, after 1 minute, the tape was stripped off by a quick peeling, no delamination or damaging of the wire was observed. In order to compare the adhesion of the cluster-assembled Pd film with a metal layer obtained by traditional evaporation technique, we deposited a Pd film on a PMMA where a gold film has been previously deposited by thermal evaporation (figure 3c). The Scotch test (figure 3d) demonstrated that the gold film was completely removed whereas the Pd film deposited directly on PMMA remained intact.

In summary we have shown that palladium clusters carried in supersonic beams can be deposited on PMMA to form patterned nanocomposite layers. The evolution of the electrical conductance with cluster coverage and TEM analysis demonstrates that clusters penetrate beneath the polymer surface already below the polymer glass-transition temperature forming a continuous and spatially defined polymer-nanoparticle nanocomposite layer. Heating the polymer favors cluster dispersion in a thicker layer below the PMMA surface. SCBD of Pd nanoparticles allows the production of metallic microstructures on PMMA characterized by a remarkably good adhesion and stability compared to metallic films obtained by thermal evaporation (despite the very weak chemical interaction of palladium with polymers [2]). This method can be used for batch fabrication [32] of metallic microstructures on flexible substrates or to impart complex physico-chemical functionalities to microfabricated polymeric devices.

**Acknowledgments:**

This work has been financially supported by Fondazione Cariplo under project "Un approccio combinatorio a materiali nanostrutturati avanzati per l'optoelettronica, la sensoristica e la catalisi". We thank M. Francolini for support in TEM characterization and L. Lorenzelli for discussions.

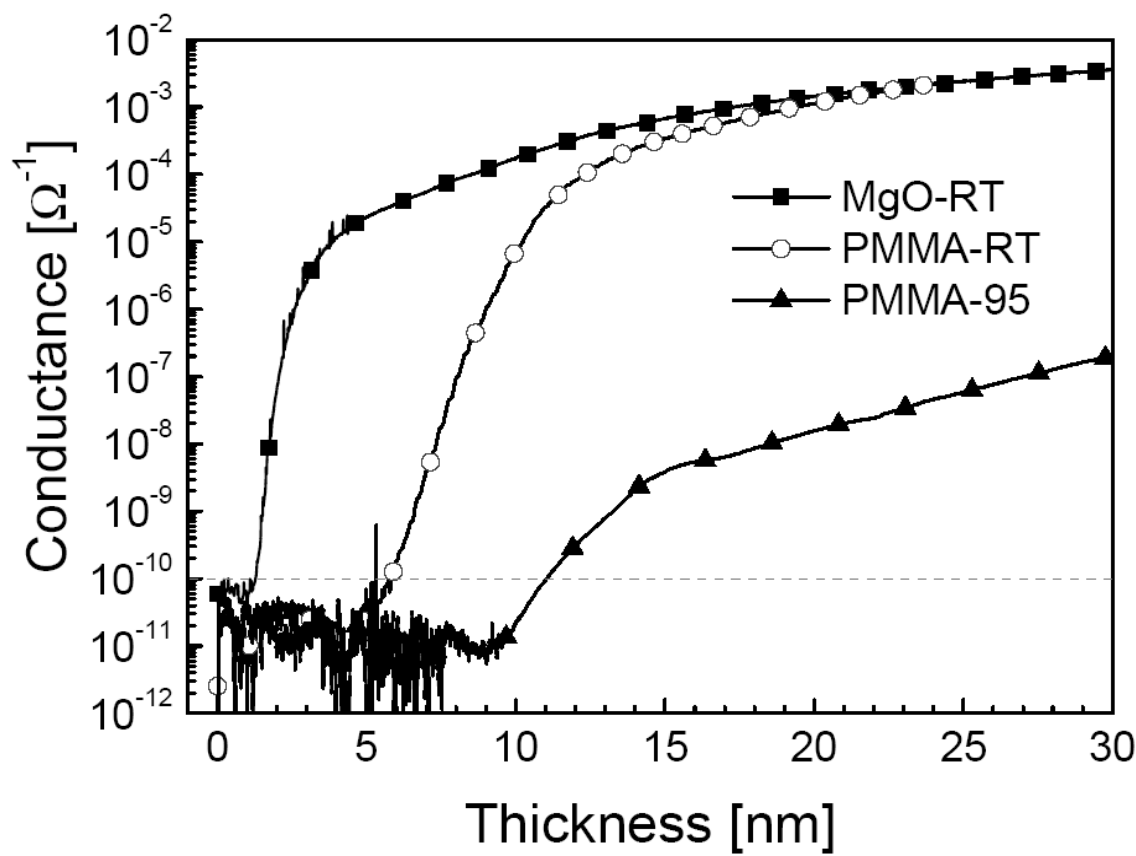

**Figure 1:** Evolution of the conductance of cluster-assembled Pd deposits as a function of their thickness (see text) on PMMA (PMMA-RT) and MgO (100) (MgO-RT) at room temperature and on PMMA at 95°C (PMMA-95). The conductance is determined in a two-probes configuration by applying a fixed potential (1.5 V) across the deposit and by measuring the current flow. Contact resistance can be neglected since the film conductance ranges in the interval of $10^{-12}$-$10^{-2}$ S. A small current is measured below the percolation threshold due to the presence of a fraction of charged clusters in the beam. The ions, landing on the gold electrodes, produce a neutralization current of about 100 pA.

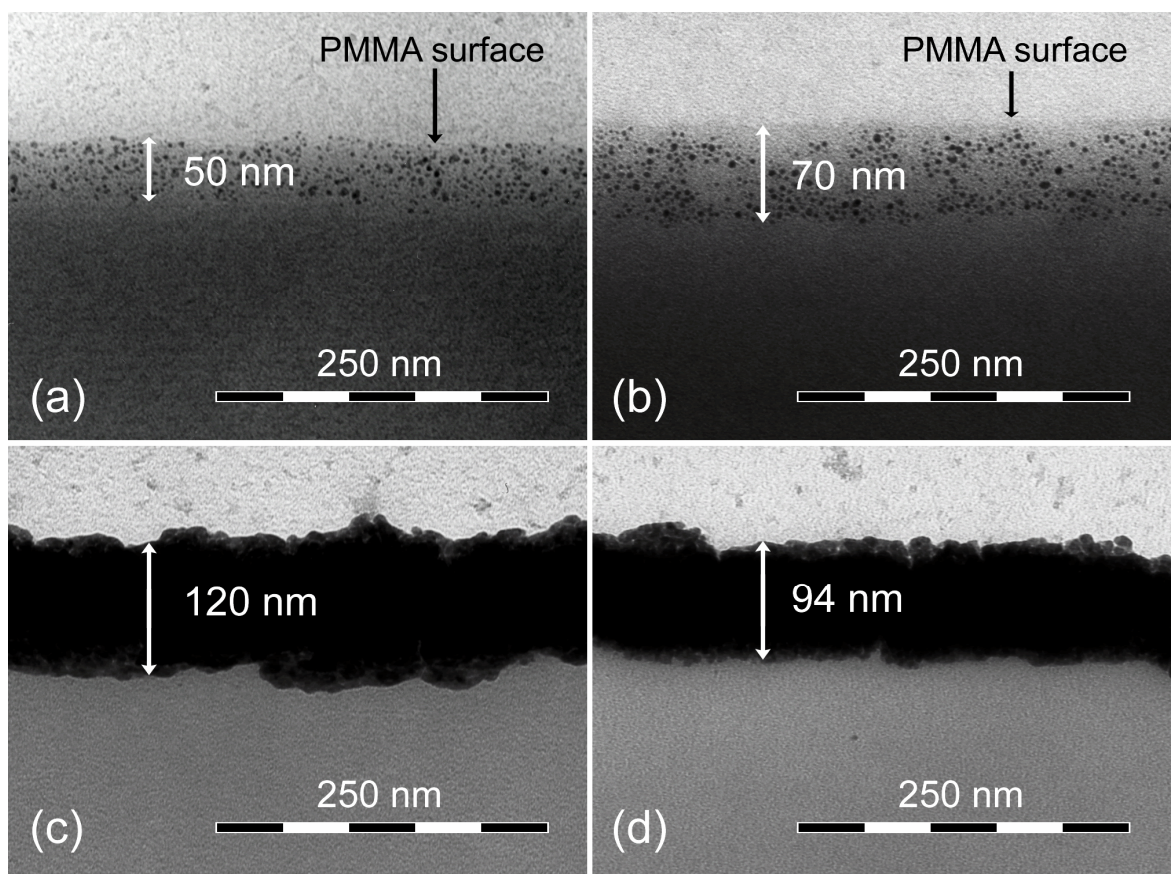

**Figure 2**: TEM micrographs (cross-sections cut by ultra-microtomy: thickness 60 nm) of Pd nanoparticles deposited on PMMA for different nominal thicknesses and substrate temperatures: **(a)** nominal thickness below 1 nm at RT; **(b)** nominal thickness below 1 nm at 95°C; **(c)** 60 nm nominal thickness at RT; **(d)** 50 nm nominal thickness at 95°C.

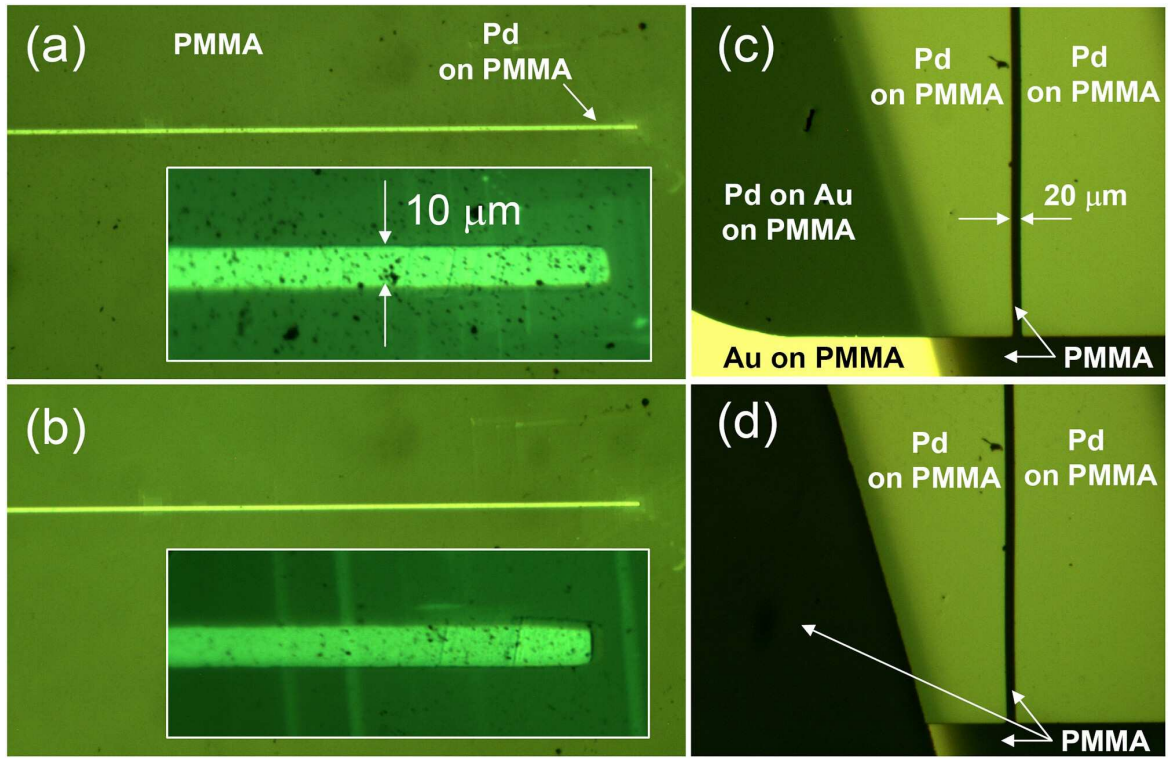

**Figure 3**: **(a)** Optical microscope images of a cluster-assembled palladium wire (2 mm long, 10 μm wide and 50 nm thick) on PMMA. The inset shows a detail of the wire. **(b)** Image of the same wire after having performed the Scotch® tape test. No damage of the wire is visible. **(c)** Optical image of a palladium nanoparticle layer deposited on a PMMA substrate. The substrate was partially coated by a previously evaporated gold film (left side in the photograph). During the Pd cluster deposition a stencil mask is used to shadow a 2 mm long and 20 μm wide trench (dark vertical line in the photo). **(d)** Image of same film after having performed the Scotch® tape test.